\documentstyle[11pt,newpasp,twoside]{article}
\index{pulsars!associations with unidentified gamma-ray sources}
\index{pulsars!millisecond}
\index{Parkes}
\index{unidentified gamma-ray sources!associations with pulsars}
\index{binaries: close}
\index{GLAST}

\def\edcomment#1{\iffalse\marginpar{\raggedright\sl#1\/}\else\relax\fi}
\reversemarginpar

\begin{document}
\title{New Pulsars Coincident With Unidentified Gamma-Ray Sources}
\author{Mallory  Roberts\footnotemark, Scott Ransom$^1$, 
Jason Hessels, Margaret
Livingstone, Cindy Tam, Victoria Kaspi}
\affil{Dept. of Physics, McGill University, Montr\'eal, QC, H3A 2T8 Canada}
\author{Fronefield Crawford\footnotemark} 
\affil{Dept. of Physics, Haverford College, Haverford, PA, 19041}

\begin{abstract}

\footnotetext{Also Dept. of Physics and Center for Space Research, MIT, 
Cambridge, MA 02139}
\footnotetext{with the Haverford College Pulsar Search Group, http://cs.haverford.edu/pulsar/}
We present the results of radio searches for pulsars within unidentified 
EGRET source error boxes. Using the Parkes multibeam system, we have
surveyed 56 sources at Galactic latitudes $|b| > 5\deg$ which do not
appear to be associated with blazars. This population has been suggested
to be distributed like the local star forming region known as the
Gould Belt, the Galactic Halo, and/or the millisecond pulsars. 
We have discovered several new pulsars in this survey, including three
new binary systems. 

\end{abstract}

\section{Introduction}

The $EGRET$ mission has left us a legacy of unidentified $\gamma$-ray sources
at low and mid-Galactic latitudes. Most of the $\sim 300$ known sources
of emission above 100 MeV remain unidentified, and most of these appear to
be associated with the Galaxy (Hartman et al. 1999). Given that the only
firmly established class of Galactic GeV $\gamma$-ray sources is 
young ($\tau < 10^6$~yr) pulsars,
it is natural to assume that many of the unidentified sources will turn out 
to be pulsars of some variety. All detections of $\gamma$-ray pulsations 
were first achieved by folding the data given an X-ray or radio ephemeris.
While blind searches of $GLAST$ data may result in 
some pulsar detections, the planned default scanning mode of observations
will complicate such searches. 
Young pulsars tend to glitch and have significant timing noise,
making coherent searches over several month long data sets problematic. For
millisecond pulsars, especially if in a short period binary, there is
little hope in  blind
searches of $\gamma$-ray data. Therefore, it is highly desirable to detect potential
$\gamma$-ray pulsars before the next generation of high-energy $\gamma$-ray
missions are launched.
 
The known $\gamma$-ray pulsars have hard $\gamma$-ray spectra. 
The unidentified sources that are bright above 1 GeV tend to lie
along the Galactic plane as would be expected for very young and
energetic pulsars. 
Potential counterparts for many of the bright,
hard, low-latitude sources have been identified, including young pulsars,
pulsar wind nebulae, and maybe some microquasars (eg. Roberts, Romani,
\& Kawai 2001). 
Unidentified sources that are faint above 1 GeV 
have a large Galactic scale height (Lamb \& Macomb 1997). 
The unidentified sources at mid-latitudes tend to have
much larger error boxes (typically $\sim 1.5^{\circ}$ diameter
95\% confidence contours) due to their softer spectrum and lower luminosity. 
This makes counterpart searching with X-ray imaging telescopes 
inefficient since it takes several pointings to cover an error
box well. Archival X-ray and radio data is also fairly sparse compared 
to what is available at low-latitudes. Therefore, the proven strategy
of identifying candidates through X-ray and radio imaging followed 
by deep pulsation searches is impractical for sources out of the plane.  

\section{A Pulsar Survey of Mid-Galactic $\gamma$-ray Sources}

The spatial distribution of unidentified $EGRET$ sources can be modelled as 
having a Gould Belt component plus a Galactic Halo component (Grenier 2002).  
Accordingly, there are at least two proposed classes of mid-latitude pulsars which might 
have detectable $\gamma$-ray emission. 
The Gould Belt is a local ($d\sim 50-300$~pc) region of recent
star formation in a rough disk inclined to the plane of the Galaxy where 
a significant number of young pulsars should have been born  (Popov et al. 2003). 
The polar-cap $\gamma$-ray emission model allows for low-luminosity 
wide-angle emission coming from outside the primary beam with a softer spectrum than 
the direct polar cap emission (Harding \& Zhang 2001). 
Such ``off-beam" $\gamma$-ray pulsars in the Gould Belt
would have been detected by $EGRET$. Many would be radio-quiet,
but a good fraction should be detectable as radio pulsars as well (Harding
et al. 2003). 
A number of the Galactic Halo sources could be millisecond pulsars (MSPs). 
The distribution of MSPs outside of globular clusters is similar to 
that of the Galactic Halo unidentified $EGRET$ sources, and
some MSPs have spin-down energies and magnetospheric potentials in the range
of the $\gamma$-ray emitting pulsars. 
There has also been a possible detection of a MSP by $EGRET$
(Kuiper et al. 2000).  

We have completed a survey of 56 unidentified $EGRET$ sources for radio
pulsations at 20~cm using the Parkes multibeam receiver. The 13 feeds of 
the receiver are placed two beam widths
apart on the sky, allowing a 4 pointing tesselation pattern to have nearly 
complete coverage of an area of the sky $\sim 1.5\deg$ across with partial
coverage out to $\sim 2\deg$, matching well the
typical mid-latitude $EGRET$ error box.  The survey pointings were 
$\sim 35$~min long, which is the same duration
as the Parkes Multibeam Galactic Survey (Manchester et al.
2001), and 8 times that of the Swinburne mid
and high-latitude surveys (Edwards et al. 2001, Jacoby et al. 2003). 
We used a sampling time of $125~\mu$s to maintain sensitivity to MSPs, and
performed full acceleration searches using the {\tt Presto} 
search software (Ransom 2001) to have sensitivity to pulsars in even 
tight binary systems. 

We used the following criteria to select our survey sources: 
$|l| > 5^{\circ}$ (so as not to overlap with the PMB plane survey); no
probable blazar counterpart as determined by Mattox, Hartman, \& Reimer (2001); 
declination $< +20^{\circ}$ for easy accessibility from the Parkes telescope;
and a 95\% error contour radius as listed in the 3rd $EGRET$ 
Catalog $\theta_{95} \la 0.75\deg$ so as to be well covered with
a single tesselation pattern. Survey observations were performed 
between June 2002 and July 2003, with confirmation observations
still ongoing. Regular timing of newly discovered pulsars commenced
in June 2003. As of this writing, we have been able to solve the orbits for 
the newly discovered binary pulsars, but we do not yet have reliable 
$\dot P$ measurements for any of the new pulsars; hence, we are 
not yet certain if they are energetically capable of being 
$\gamma$-ray sources. 

A total of 13 pulsars have been detected and confirmed so far in the survey. 
Six of the pulsars are new discoveries, and two 
pulsars in the survey area are listed in the ATNF catalog
(http://www.atnf.csiro.au/research/pulsar/psrcat/) 
but were not redetected. Interestingly, only one pulsar
(the 669 day binary PSR J0407+1607) was detected further
than $\sim 50\deg$ from the Galactic center (most of the
pulsars are within $\sim 30\deg$), despite more than
half of the sources surveyed being outside this region. 
None of the pulsars are potential Gould Belt sources. 

Only 3 new isolated pulsars were discovered, much fewer than were expected. 
The reason for this is probably man-made radio frequency interference (RFI). 
Although RFI masking was implemented and the Fourier spectra de-reddened, 
it was obvious from the folding of candidates that, for long periods 
($\ga 200$~ms) and low trial dispersion measures ($DM \la 50$), the
$P-\dot P $ plane was filled with noise well above our sensitivity 
threshold. Unfortunately, this region of parameter space is exactly where 
we expect Gould Belt pulsars to lie. In order to test that the
lack of new slow pulsar detections was not due to some unknown
bias in the {\tt Presto} software system, the data was searched 
independently  for slow ($> 4$~ms) pulsars using 
the {\tt Sigproc/Seek} (D. Lorimer, http://www.jb.man.ac.uk/~drl/) software 
package. No new pulsars were discovered in the reanalysis. 
Additional reanalysis using much stricter RFI excision is planned.  

\section{Binary Pulsars Discovered in $EGRET$ Error Boxes}

The other three new pulsars appear to be recycled pulsars
with low mass companions. A fourth binary pulsar,
PSR J0407+1607, was detected which had previously been discovered at Arecibo 
(Lorimer et al. in preparation). This is the most productive 
survey yet for the number of binary pulsars detected per square
degree of sky outside of globular clusters. A fifth
binary pulsar is known within the survey area that was not redetected. 
4 of these 5 binary pulsars are within $30\deg$ of the galactic center,
suggesting a deep survey of this region for binary
pulsars would be very productive.

Remarkably, two of the new binary systems were discovered towards
the same $EGRET$ source, 3EG J1616$-$2221. PSR J1614$-$22 is a 3.15~ms
pulsar in an 8.7 day orbit around a white dwarf companion. PSR J1614$-$23
is a 33.5~ms pulsar in a 3.15 day orbit with a minimum companion mass of
only $0.08 M_{\sun}$. A 350 MHz observation of PSR J1614$-$23 with the GBT 
shows that the base of the
pulse profile is very broad, perhaps suggesting a small inclination angle.
The third new binary pulsar, PSR J1744$-$39, 
is at a Galactic latitude of $b=-5.3$, at
the edge of the Parkes Multibeam Survey region, and was in fact
independently discovered in the recent reanalysis of the PMB survey
data (Andrew Lyne, private communication). PSR J1744$-$39 is 
a 172~ms pulsar in a 4.6 hr. binary system  also with a 
minimum companion mass of $0.08 M_{\sun}$. This pulsar is notable for
turning on and off on timescales of $\sim 100$~s, despite having a long
period and a fairly high dispersion measure ($DM=147.5\, {\rm pc}\, 
{\rm cm}^{-2}$). Continued timing of these sources will shortly determine
whether they have the power to be the source of the observed $\gamma$-rays.

\begin{table}
\begin{center}
\caption{New Pulsars discovered in Survey}
\begin{tabular}{lcccc}
\tableline
Pulsar  & $P$ & $D$\tablenotemark{a} & $P_B$\tablenotemark{b} & $M_c$
\tablenotemark{c} \\
& s & kpc & d & $M_{\sun}$ \\
\tableline
J1614$-$22 & 0.003151 &  1.3 & 8.68 & 0.4 \\
J1614$-$23 & 0.03350 & 1.9 & 3.15 & 0.08  \\
J1632$-$10 &  0.7176  & $ > 50$ & - & -  \\
J1725$-$07 & 0.2399 &  1.7 & - & -  \\
J1744$-$39 & 0.1724& 3.1 & 0.19 & 0.08 \\
J1800$-$01 & 0.7832 &  1.6 & - & -  \\
\tableline
\tableline
\tablenotetext{a}{Distance from the NE2001 dispersion measure model of
Cordes and Lazio}
\tablenotetext{b}{Binary orbital period}
\tablenotetext{c}{Minimum companion mass}
\end{tabular}
\end{center}
\end{table}

We have begun a northern extension of this survey using the
GMRT at 610 MHz. Hopefully, the lower frequency will help us maintain 
sensitivity to slower, nearby pulsars, allowing us to put meaningful
limits on the number of Gould Belt pulsars.

\end{document}